\documentclass[letterpaper, 10 pt, conference]{ieeeconf}
\IEEEoverridecommandlockouts
\usepackage{amsfonts}
\usepackage{graphicx}
\usepackage{stfloats}
\usepackage{amsmath}
{
      \newtheorem{assumption}{Assumption}
}

\usepackage{physics}
\newcommand\addtag{\refstepcounter{equation}\tag{\theequation}}

\newtheorem{theorem}{Theorem}

\newtheorem{Proposition}{Proposition}
\title{\LARGE \bf Parametric model order reduction for \\large-scale and complex thermal systems}
\author{Daming Lou  and Siep Weiland$^{1}$
\thanks{$^{1}$Daming Lou, Siep Weiland are with the Department of Electrical Engineering, Eindhoven University of Technology, Eindhoven, The Netherlands. Emails: {\tt\small \{D.Lou, S.Weiland\}@tue.nl}}
}

\begin{document}

\maketitle
\thispagestyle{plain}
\pagestyle{plain}
\begin{abstract}
In this paper, a parametric model order reduction (pMOR) technique is proposed to find a simplified system representation of a large-scale and complex thermal system. 
The main principle behind this technique is that any change of the physical parameters in the high-fidelity model can be updated directly in the simplified model. 
For deriving the parametric reduced model, a Krylov subspace method is employed which yields the relevant subspaces of the projected state. With the help of the projection operator, first moments of the low-rank model are set identical to the correspondent moments of the original model. Additionally, a prior upper bound of the error induced by the approximation is derived.
\end{abstract}
\

\IEEEpeerreviewmaketitle

\section{Introduction}
Numerical simulations of complex dynamical systems are an indispensable tool in studying thermodynamic phenomena. However, for complex thermal systems where ultra-high precision simulations are required, the finite element method (FEM) commonly yields large-scale models. These models demand considerable computational resources. Therefore, model order reduction techniques are employed 
to reduce the computational complexity by replacing the high-order dynamic model with a low-order one. For the class of  linear time invariant (LTI) systems, many reduction 
techniques have reached a relatively high level of maturity \cite{Antoulas}. For systems with uncertain or time-varying parameters there is a persistent need of novel reduction techniques. 
Since creating a new reduced model for every parameter value is inefficient and computationally costly, there is a strong need for parametric model order reduction (pMOR) techniques as introduced by \cite{PeterBenner}. 
We distinguish among reduction techniques in which both the state dimension and parameter vector dimension are reduced, and techniques in which state dimension is reduced while the parameter vector keeps its physical relevance. In addition, we distinguish among time-varying and time-invariant parametric dependence of high-fidelity models. In either of these cases, the objective is to find a low-cost, but accurate, parameterized model.\\
\indent In the last decades, model order reduction (MOR) techniques based on Pad\'e approximation \cite{antoulas1986recursiveness} have been used as powerful tools for large-scale system simulation, in particular for very-large-scale system integration (VLSI) simulation \cite{Gugercin} and microelectromechanical (MEMS) simulation tasks \cite{panzer2010parametric}. The first algorithm was an asymptotic waveform evaluation in \cite{AWE}, which uses explicit moment-matching. For a single-input single-output (SISO) system, the Arnoldi algorithm, also called the one-sided moment-matching technique \cite{One-sided} was proposed to match $r$ moments of a transfer function by a $r$th-order approximation. Meanwhile, two-sided moment matching, known as Lanczos method \cite{PVL}, was introduced to perform matching of $2r$ moments by a $r$th-order approximations. Then the block Arnoldi \cite{BlockOne-sided} and block Lanczos method \cite{BlockTwo-sided} were proposed to solve the same question of  multi-input multi-output (MIMO) systems.
However, for large-scale systems which are parameter dependent, 
the aforementioned methods are not suitable anymore for obtaining a low-order model while maintaining the physical interpretation after the MOR procedure. Therefore, there is a need to develop such MOR techniques that allows the reduction of parameter-varying systems. Hence, a new branch of MOR is known as parametric model order reduction (pMOR).

The first work of pMOR using Krylov subspace methods was introduced in \cite{1stPMOR}, where the state evaluation matrix linearly depends on a single parameter. In consideration of the physical systems that can be described with multiple parameters, an extensive pMOR method which matches the coefficients of multivariate Taylor series is generalized in \cite{PMORMM}. This method 
replaces the product of complex independent variables $s$ and the parameter $p$ in the transfer function $G(s,p)$ by a set of redefined parameters, where the moments of the reduced model are equivalent to the corresponding moments of the original model with respect to the new expansion points. Contrary to singular value decomposition (SVD) based truncation, the pMOR techniques based on Krylov subspaces have no prior error bound or global error bound.  


In this paper we explore MOR techniques for parametric models with the purpose to match multiple moments of the transfer function. We introduce a separation between the physical parameters $p$ and the frequency points $s$ such that, unlike previous work, the physical interpretation of the physical parameters in the reduced model is maintained. Also, simulation results demonstrate the benefits of this separation. Moreover, an upper bound of the approximation error between the reduced model and the original model is derived, which allows prior estimating the misfit without calculating the exact reduced model.

The remainder of the paper is structured as follows: In Section II, a condensed problem description for parametric model order reduction is presented. The proposed method for parametric model order reduction with moment matching of multiple frequencies and multiple parameters is introduced in Section III. The analytic expression of the upper bound of this method is given in section IV. Simulation results are shown in Section V. Finally, Section VI concludes this work.

\section{Problem Description}
The main goal of the reduction techniques proposed in this paper is to preserve parameters in the system as symbolic quantities in the reduced order model. In fact, there are many key applications where this parametric model reduction technique is of crucial interest. These include design optimizations (where the parameter vector represents design parameters) in which the design-loop can substantially benefit from simplified parametric models, calibration of parameters through simplified models or control design for parameter dependent models. In either of these cases, a change in parameters does not require to repeat the reduction procedure, but simply the evaluation of the parametric model. The aim is to infer a reduced order model with smaller state dimension, but which remains explicit in the parameter $p$. 

Consider a linear parameter-varying model with parameter ${p}(t) = [ p_{1}(t),\cdots,p_{\ell}(t) ]^{T} \in \mathbb{R}^{\ell}$ described as 
\begin{equation} \label{equ: ess}
\Sigma({p}):= \left\{ \begin{array}{ll}
 E({p}) \dot{x}(t) = A({p})x(t)+B({p})u(t) \\
  \hspace{0.75cm}  y (t) = C({p})x(t)
\end{array} \right.
\end{equation}
where $x(t)\in \mathbb{R}^n$, $u(t)\in \mathbb{R}^m$ and $y(t) \in \mathbb{R}^q$ denote, respectively, the state vector, the input and the output. Here $n$ is the cardinality of the mesh of a spatially distributed configuration space $\Omega$ (usually of dimension $3$). The state-space matrices are functions $A: \mathbb{R}^{\ell} \rightarrow \mathbb{R}^{n \times n}$, $E: \mathbb{R}^{\ell} \rightarrow \mathbb{R}^{n \times n}$ etc. 
Before explaining the model reduction, we first state several assumptions.

\begin{assumption}\label{assumption: 1}
i) $[E(p),A(p),B(p),C(p)]$ system matrices are affine functions of $p\in \mathbb{R}^\ell$. e.g., $E({p}) = E_0+ \Sigma_{i =1}^{\ell} E_ip_i $
 ii) ${p}\in \mathbb{P} \subset \mathbb{R}^{\ell}$ is not varying with time. 
iii) For all $p \in \mathbb{P}$, all generalized eigenvalues of $(\lambda E(p)-A(p))$ have negative real parts.
\end{assumption}

With the time-invariant $p$, the transfer function of (\ref{equ: ess}) is meaningful and is given as follows:
\[\addtag \label{equ: essTF}
G(s,p) = C(p)\big[sE(p)-A(p) \big]^{-1}B(p).
\]


\indent We consider a projection-based method for generating the reduced order parametric model. That is, let $V\in \mathbb{R}^{n \times r}$ and $W \in \mathbb{R}^{n \times r}$ be full rank matrices with $r \ll n$ and let
\[\label{equ: projection}\addtag
  x = Vx_r,\quad V \in \mathbb{R}^{n \times r}\quad \text{with} \quad {x_r} \in \mathbb{R}^{r}.
\] 

Define the projection spaces
\[\label{eq: projection base} \addtag
\mathcal{V}: = \text{im}(V); \quad \mathcal{W}: = \text{im}(W);
\]
i.e. two $r$-dimensional subspaces of $\mathbb{R}^n$. Let $\Pi_{\mathcal{V}}$ and $\Pi_{\mathcal{W}}$ denote the (canonical) projections of $\mathbb{R}^n $ onto $\mathcal{V}$ and $\mathcal{W}$, respectively. That is $\Pi_{\mathcal{V}} = V(V^TV)^{-1}V^T$ and $\Pi_{\mathcal{W}} = W(W^TW)^{-1}W^T$ and consider the system matrices of the reduced order parametric model defined by
\begin{eqnarray}\label{eq: dress}
      E_{r}({p})= W^{T}E({p})V,\quad A_{r}({p})= W^{T}A({p})V \nonumber  \\
      B_{r}({p})= W^{T}B({p}),\quad C_{r}({p})=C({p})V.
  \end{eqnarray}
and the state space representation of the reduced model is then given by
\begin{equation}\label{eq: rDSS}
  \Sigma_{r}({p}):= \left\{ \begin{array}{ll}
 E_{r}({p}) \dot{x}_{r}(t) = A_{r}({p})x_{r}(t)+B_{r}({p})u(t) \\
  \hspace{1.04cm}  y(t)  = C_{r}({p})x_{r}(t)
\end{array} \right.
\end{equation}
The transfer function of the parametric reduced model is
\[\addtag \label{equ: rDSSTF}
G_r(s,p) = C_r(p)\big[sE_r(p)-A_r(p)\big]^{-1}B_r(p)
\]
If $W = V$, this is called an (ordinary) Galerkin projection. If $W \neq V$, this is a Petrov-Galerkin projection. 

\section{Parametric Model order reduction}
In this section, the moment-matching method is elaborated first for a non-parametric model. 
In the second part, we present a parametric moment-matching method which separates the frequency variable $s$ and the parameter $p$. 

\subsection{Moment-matching}
Suppose that a non-parametric model is given
\begin{equation} \label{eq: sisosys}
\left\{ \begin{array}{ll}
 E \dot{x}(t) = Ax(t)+Bu(t) \\
  \hspace{0.3cm}  y (t) = Cx(t)
\end{array} \right.
\end{equation}
with $x(t) \in \mathbb{R}^n, u(t)\in \mathbb{R}^m$ and $y(t) \in \mathbb{R}^q$.
Now, let $s_0 \in \mathbb{C}$ be fixed and expand the Taylor series of the transfer function $G(s先)$ of (\ref{eq: sisosys}) at $s_0$. That is
\begin{eqnarray}\label{equ: taylor expansion}
    G(s)&=& C((s_0+\sigma)E-A)^{-1}B \nonumber \\
      &=& \sum \limits^\infty_{i=0}\underbrace{C\Bigl[-(s_0{ E}-{A})^{-1}E\Bigr]^i(s_0{ E}-{ A})^{-1}B}_{:= m_i(s_0)} \, \sigma^i \hspace{0.4cm}
\end{eqnarray}
where $s = s_0 +\sigma$ and the complex matrices $m_i(s_0)$ are known as the \textit{moments} of $G(s)$ at expansion point $s_0$. Define the block Krylov subspace of order $r> 0 $ by
\begin{eqnarray}
 \mathcal{K}_r(M,f) &=& \text{colspan}\{f,Mf,\cdots,M^{r-1}f\}  \label{eq: V_proj}
\end{eqnarray}
where
\begin{eqnarray}\label{eq:InputKry}
 f &=& (s_0E-A)^{-1}B \in \mathbb{R}^{n\times m}\\ \nonumber
 M &=&(s_0E-A)^{-1}E \in \mathbb{R}^{n \times n} \nonumber.
 \end{eqnarray}

Then $\mathcal{K}_r(M,f)$ has dimension $r$ and the columns of the matrix $V$ in (\ref{equ: projection}) are constructed as a basis of $\mathcal{K}_r(M,f)$. Matrix $W$ is chosen such that $W^TAV$ is nonsingular. Then the transfer function $ G_r(s)$ of the resulting projection-based reduced model is obtained by (\ref{equ: rDSSTF})
and the reduced system satisfies the following condition
\[ \addtag
m_{r,j}(s_0) = m_j(s_0), \quad j = 0,\dots,r-1
\]
Here, the $m_{r,j}(s_0)$ and $m_j(s_0)$ denote the $j$th moment of the reduced model at $s_0$ and the $j$th moment of the original model, respectively.


The aforementioned method describes the one-sided moment matching. By choosing the columns of $W$ as a basis of the Krylov subspace
\[ \addtag \label{eq: W_proj}
\mathcal{K}_r(M_2,l) = \text{colspan}\{l,M_2l,\cdots,M_2^{r-1}l\}
\]
where
\begin{eqnarray}
l &=&  (s_0E-A)^{-T}C^T \in \mathbb{R}^{n \times q}\nonumber \\
M_2 &=& (s_0E-A)^{-T}E^T \in \mathbb{R}^{n \times n}
\end{eqnarray}
another $r$ moments can be matched. The proof can be found in \cite{Proof_MomentMatching}.
\theorem \textit{Let the projection matrices $V$ and $W$ be such that $\mathcal{V}$ and $\mathcal{W}$ in (\ref{eq: projection base}) coincide with (\ref{eq: V_proj}) and (\ref{eq: W_proj}), respectively. Then, the first $2r$ moments $m_i(s_0)$ and $m_{r,i}(s_0)$, $i= 0,\cdots, 2r-1$ of $G(s)$ and $G_r(s)$ defined in (\ref{equ: essTF}) and (\ref{equ: rDSSTF}) coincide.}


It is worth to mention that the projection matrices $V$ and $W$ can have complex numbers with complex expansion points, as a result, the system matrices of the reduced model may contain complex numbers which is often physically unreasonable. One solution is to take $s_0$ real or to have a pair of complex conjugate expansion points $(s_0,\bar{s_0})$ and the reduced model results in real-valued matrices.

\subsection{Multi-parameter and multi-frequency moment-matching}
The main ingredient of the multi-parameter and multi-frequency moment-matching method is to expand the transfer function into a Taylor series at both a desired frequency and a desired parameter $(s_0,p_0)$. This work has been generalized in \cite{PMORMM}. 

Using (\textit{Assumption \ref{assumption: 1}}), we obtain
\begin{eqnarray} \label{eq: multi-ess}
  G(s,{p}) 
   &=& C({p})[sE_0+\Sigma_{i =1}^{\ell} sE_ip_i- A({p})]^{-1}B({p})\hspace{0.8cm}
\end{eqnarray}
where the product of the frequency $s$ and the parameter $p_i$ can be redefined as $\hat{s}_i= s\cdot p_i$. This has been proposed in \cite{PMORMM}. 
Potentially, combining the $s$ with $p_i$ as a single variable $\hat{s}_i$ can decrease the complexity of the power series. In such a way, the computational burden of the projection matrices $V$ and $W$ is reduced. However, with such conversion the reduced model matches the first moments of the original model with respect to $\hat{s}_i$ and the physical interpretation of the parameters $p$ in the reduced model are no longer preserved. Moreover, the Taylor expansion of the new parametric transfer function includes terms such as $(\frac{\partial^2 G}{\partial \hat{s}_i \partial\hat{s}_{i+1}})$ which are partial derivatives of composed functions rather than the original one. 

We define two sets $\mathcal{S}:=\{s_1,\dots,s_k\} \subset \mathbb{C}$ and $\mathcal{P}:=\{p_1,\dots,p_{\ell}\}\subset \mathbb{R}$ which denote the frequency variables and design parameters of interest, respectively. The projection matrices $V$ and $W$ can be constructed along these two sets.

The following details the parametric moments which separate the frequency $s$ from the parameters $p$. The Taylor series of (\ref{eq: multi-ess}) around the expansion point $(s_i, p_j)\in \mathcal{S} \times \mathcal{P}$ is given below
\begin{align}\label{eq: TaylorMM}
G(s,{p})& = G{({s}_i,{p}_j)}  \nonumber\\
&+ \frac{\partial G}{\partial s}(s_i,p_j) (s-s_i) + \frac{\partial G}{\partial p}(s_i,p_j)(p-p_j)\nonumber \\
&+ \frac{1}{2}\frac{\partial^2 G}{\partial^2 s}{({s}_i,{p}_j)}(s-s_i)^2 +\frac{1}{2}\frac{\partial^2 G}{\partial^2 p}({s_i},p_{j})(p-p_j)^2 \nonumber \\
&+ \frac{1}{2}\bigg[\frac{\partial^2 G}{\partial s\partial p}+\frac{\partial^2 G}{\partial p\partial s}  \bigg](s_i,p_j)(s-s_i)(p-p_j) \nonumber \\
 &+ \cdots
\end{align}

\begin{figure}[h]
  \centering
  \includegraphics[scale = 0.28]{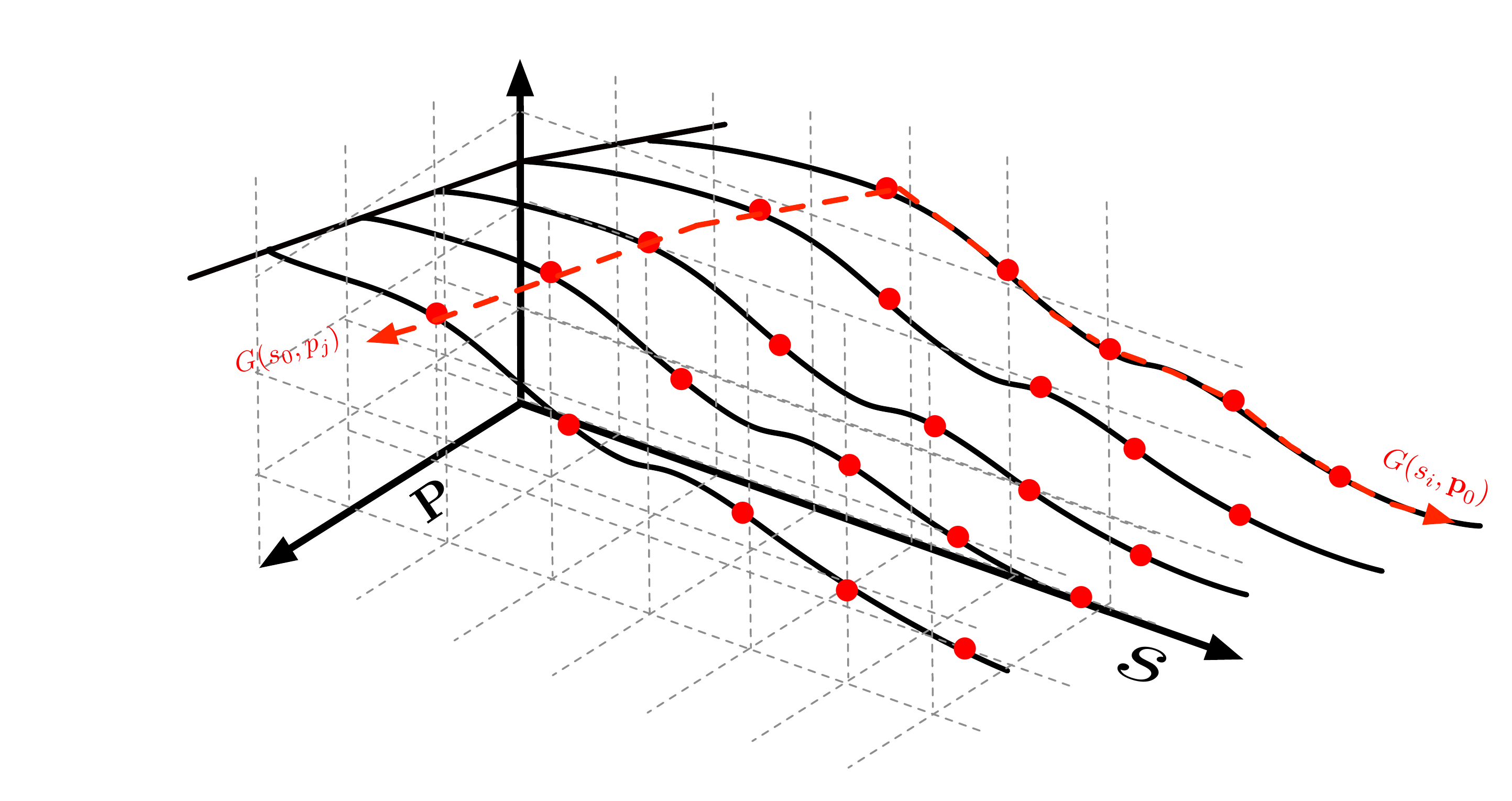}
  \caption{Multi-frequency and multi-parameter expansion}\label{Fig: MM_multi}
\end{figure}

Here we present the Taylor series for a single point. Applying this for $\mathcal{S}\times \mathcal{P}$ yields multiple expansion points as depicted in Figure \ref{Fig: MM_multi}. The parametric moments for multiple expansion points are defined as
\begin{eqnarray}
m_{i}(s,{p}): &=& C(p)\big[-(sE(p)-A(p))E(p)\big]^{i}\cdot \nonumber\\
            &&  (sE(p)-A(p))^{-1}B(p)
\end{eqnarray}
where $(s,p) \in \mathcal{S}\times \mathcal{P} $.

The parametric projection matrices $V$ and $W$ are now defined as follows.
\begin{align}
\mathcal{V}  &=  \mathcal{K}_r(\boldsymbol{M}_1,F) \label{eq: V_proj2}  \\
\mathcal{W}  &=  \mathcal{K}_r(\boldsymbol{M}_2,L)  \label{eq: W_proj2}	
\end{align}
where
\begin{align}
&F  =  \text{rowspan}\Big\{(s_i E(p_j)- A(p_j)) ^{-1} B(p_j) \Big\} _{i=1,j=1}^{k,\ell}  \nonumber \\
&L  = \text{rowspan}\Big\{ (s_i E(p_j)-A(p_j))^{-T}C(p_j)^T \Big\}_{i=1,j=1}^{k,\ell} \nonumber \\
&\boldsymbol{M}_1  =  \text{rowspan}  \{(s_iE(p_j)-A(p_j))^{-1}E(p_j)\}_{i=1,j=1}^{k,\ell} \nonumber \\
&\boldsymbol{M}_2  =   \text{rowspan}\{ (s_iE(p_j)-A(p_j))^{-T}E(p_j)^T\} _{i=1,j=1}^{k,\ell}.
\end{align}

As in (\ref{equ: rDSSTF}) with (V,W) satisfies (\ref{eq: projection base}), this yields a reduced order model whose transfer function
\[ \label{eq: multi-ress}\addtag
  G_r(s,{p}) 
   = C_r({p})[sE_{(r,0)}+\Sigma_{i =1}^{\ell} sE_{(r,i)}p_i- A_r({p})]^{-1}B_r({p})
\]
coincides in as many moments of the Taylor series  of the original model about $(s,p)$ as possible for a given order of the reduced model. That is, we claim that
\[\addtag \label{eq: paraMM}
m_{r,i}(s,p) = m_i(s,p)
\]
for all $(s, p) \in (\mathcal{S}\times \mathcal{P})$ and $i = 0,\cdots,2r-1$.
\theorem \textit{Let the projection matrices $V$ and $W$ be such that $\mathcal{V}$ and $\mathcal{W}$ in (\ref{eq: projection base}) coincide with (\ref{eq: V_proj2}) and (\ref{eq: W_proj2}), respectively. Then, the first $2r$ moments $m_{r,i}(s,p)$ and $m_i(s,p)$, $i = 0,\cdots,2r-1$ of $G(s,p)$ and $G_r(s,p)$ defined in (\ref{eq: TaylorMM}) and (\ref{eq: multi-ress}) coincide for all $(s,p) \in \mathcal{S}\times \mathcal{P}$.}

%


\section{Error estimation}
Guarantees on the fidelity of the reduced model remain an important question for the previously described moment matching procedure. One major issue is establishing
a prior estimation of the approximation error. In the following we will derive an error estimate between $G(s,p)$ and $G_r(s,p)$ which provide a local accuracy at points $(s_0,p_0)\in \mathcal{S}\times\mathcal{P}$ and their neighborhood. That is, the error estimate of $G(s,p)-G_r(s,p)$ will have local validity in $(s_0,p_0)$ and is based on the truncation order of the Taylor series.
%
%
%

\subsection{The remainder in Taylor series}
Let $f: \mathbb{R} \rightarrow \mathbb{R}$ 
be differentiable on the open interval $(a,b)$. Then a point $c$ with $a<c<b $ exists such that
\[
f^{\prime}(c) = \frac{f(b)-f(a)}{b-a} \addtag.
\]

The above equation is the Mean Value Theorem (MVT) which is a fundamental theorem of calculus. Taylor formula can be viewed as a generalization of MVT. In particular, let $f$ be a function such that $f$ and its first $N+1$ derivatives are continuous at $a$, then the Taylor series is
\[
f(x) = f(a) + \frac{f^{\prime} (a)}{1!}(x-a) + 
\cdots + \frac{f^{(N)} (a)}{N!}(x-a)^N+ \cdots \addtag
\]
and the $N$th-degree Taylor polynomial is obtained
\begin{eqnarray}\label{eq: SingleTaylorPoly}
f_N(x) 
       = f(a)+ \sum_{i= 1}^{N}\frac{f^{i}(a)}{i!}(x-a)^{i}
\end{eqnarray}

Thus we can write Taylor's theorem
\[ \label{eq: Remaind123}\addtag
f(x) = f_N(x) + R_N(x)
\]
where $R_N(x)$ is the remainder term which quantifies the difference between the Taylor series and the $N$th-degree Taylor polynomial in terms of the magnitude of the $(N+1)$st derivative of $f$.

\begin{theorem}\textit{\label{theorem: remainder}
  Suppose that $f: \mathbb{R} \rightarrow \mathbb{R}$ is defined on an interval $I$ that has $a$ in its interior and $f^{(N+1)}$ exists on the same interval for $N \in \mathbb{Z}_{+}$. Then for each $x \neq a, x\in I$, there is a $\xi$ between $x$ and $a$ so that
\begin{eqnarray}\label{eq: Remaind}
R_N(x)= \frac{f^{(N+1)}(\xi)}{(N+1)!}(x-a)^{N+1}.
\end{eqnarray}
}\end{theorem}

Thus, there exists a $\xi \in I$, with either $x\leq \xi \leq a$ or $a\leq \xi \leq x$ such that the remainder $R_N(x)$ in (\ref{eq: Remaind123}) satisfies (\ref{eq: Remaind}).

A key observation is that when $N = 1$, this reduces to the ordinary MVT. 
Following this strategy, the key is to observe the generalization of Rolle's theorem.

\begin{Proposition}\label{theorem: rolle's}
Suppose $f:\mathbb{R} \rightarrow \mathbb{R} $ is a function on interval $I$. Assume that $f$ is $N\in \mathbb{Z}^+$ times differentiable on the interval $(a,b) \in I$, and that $f,f^{\prime},\cdots,f^{N-1}$ all extend continuously to the interval $[a,b]$. If in addition
\[\addtag
 f(a) = f^{\prime}(a) =\dots = f^{N-1}(a) = 0=f(b)
\]

Then there exists such $c\in (a,b) $ that
\[\addtag f^{N}(c) = 0.
\]
\end{Proposition}

The \textit{Theorem \ref{theorem: remainder}} can be proved by applying \textit{Proposition \ref{theorem: rolle's}} and the  proof is omitted due to lack of space.
%
%
%

Furthermore, we can find the upper bound of the remainder term in \textit{Theorem \ref{theorem: remainder}} using  Taylor's inequality.

\begin{theorem}\label{theore: Taylor inequality} \textit{ Under the conditions of \textit{Theorem \ref{theorem: remainder}}, let $I_a \subseteq I$ be any interval that has $a$ in its interior. We define
\[\addtag \label{eq: InequalitySISO}
\mathcal{M}:= \sup_{\xi \in I_a} \left\vert \frac{f^{(N+1)}(\xi)}{(N+1)!}\right\vert
\]
Then
\[\addtag \label{eq: UpperboundSISO}
|f(x)-f_N(x)| = |R_N(x)|\leq \mathcal{M}|x-a|^{N+1}
\]
holds for all $x\in I_a$.
}\end{theorem}

Hence an upper bound of the error between the $N$th degree Taylor approximation $f_N$ and $f$ is obtained and is a prior bound, only depending on $f$ itself and the considered interval $I$. In the following, we generalize \textit{Theorem \ref{theore: Taylor inequality}} to multivariable function $f: \mathbb{C}^{\ell} \rightarrow \mathbb{C}^{m\times p}$ for MIMO systems with $\ell$ expansion points, $m$ inputs and $q$ outputs.

Suppose $f: \mathbb{C}^{\ell} \rightarrow \mathbb{C}^{m\times p}$ is $N+1$ times differentiable on a set $I\subset \mathbb{C}^{\ell}$ that contains a vector ${a}\in \mathbb{C}^{\ell}$ in its interior. Let $I_{{a}} \subset I $ be a closed and bounded subset of $I$. Then the truncated $N$th order Taylor polynomial around ${a}\in \mathbb{C}^{\ell}$ at order $N$ is
\begin{eqnarray}\label{eq: TaylorPolyUpper}
	f_N({x}) &=& f({a}) + \sum_{i = 1}^{\ell} \frac{\delta^i }{\delta x_i}f(x)\bigg\vert_{{x} = {a}} (x_i-a_i)+\cdots \nonumber\\
	 &+&	\frac{\delta^{n_1}\delta^{n_2}\cdots\delta^{n_\ell}}{\delta x_1^{n_1}\delta x_2^{n_2}\cdots \delta x_\ell^{n_\ell}}f(x) \bigg\vert_{{x} = {a}}(x_1-a_1)^{n_1} \nonumber \\
	 && \cdots(x_{\ell-1}-a_{\ell-1})^{n_{\ell-1}} (x_\ell-a_\ell)^{n_\ell}
	 \end{eqnarray}
where $\sum_{i=1}^{\ell}n_i=N$ and where $n=(n_1,\ldots, n_{\ell})$ is a multi-index of non-negative integers $n_i$

For any such multi-index $k = (k_1,\cdots,k_\ell)$ we let $|k|= \sum_{i}^{\ell}k_{i}$ be its cardinality. To simplify notation in (\ref{eq: TaylorPolyUpper}  ), the $k$th partial derivative of $f$ at a point $a \in \mathbb{C}^{\ell}$ is denoted by 
\[\addtag \label{eq: SimpleNotation}
[D^{|k|}f]_{k}:= \frac{\partial^{k_1}\cdots \partial^{k_\ell}}{\partial x_1^{k_1}\cdots \partial x_\ell^{k_\ell}}f(x)\bigg\vert_{x=a}.
\]

When ranging over all multi-indices $k$ with $|k|= K$, $[D^{|k|}f]_k$ can be interpreted as the coefficients of a order $|k|$ tensor (i.e. multi-linear functional)
\[\addtag \label{eq: tensorDefine}
 [D^{|k|}f]:\underbrace{\mathbb{C}^{\ell} \times \mathbb{C}^{\ell} \times \cdots \times  \mathbb{C}^{\ell} }_{|k| \text{ copies}} \rightarrow \mathbb{C}
\]
that has (\ref{eq: SimpleNotation}) in its elements.

Thus, $D^0f(a)$ is a number. $D^1f= \nabla f(a)$ is the gradient of $f$ at the point $a$. $D^{2}f = \nabla^2 f(a)$ is the Hessian matrix of $f$ at the point $a$ and
\[\addtag \label{eq: TensorWithK}
[D^N f](v_1,\cdots,v_N): = \sum_{|k|=0}^{N} [D^N f]_{k}(v_1 \otimes \cdots \otimes v_N)
\]

With the tensor definition of $[D^0f],[D^1f],\cdots,[D^Nf]$, the $N$th order multivariable of $f$ at $a\in \mathbb{C}^{\ell}$ (\ref{eq: TaylorPolyUpper}) can be rewritten as
\begin{eqnarray}\label{eq: simpliedTaylorPoly}
f_N(x) &=& [D^0f]+[D^1f]\big([x-a]\big) + \cdots  \nonumber \\
&&+\frac{1}{N!}[D^{|N|}f]\bigg(\underbrace{ [x-a],\cdots,[x-a]}_{N\text{ copies}}\bigg).
\end{eqnarray}

Define the error by the residual which has the similar form as the single variable one (\ref{eq: Remaind123})
\[\addtag
R_N(x): = f(x)-f_N(x)
\]
and also define $k = (k_1,\cdots,k_\ell)$ 
\[
(x-a)^k:= \prod _{j=1}^{\ell}(x_j-a_j)^{k_j}.
\]

Let, as (\ref{eq: InequalitySISO})
\[\addtag\label{eq: upper2}
\mathcal{M}:= \sup_{\xi\in I_a}\bigg\| \frac{D^{(N+1)}f(\xi)}{(N+1)!}\bigg\|
\]
where the norm is the Frobenius norm of a multi-linear operator and the supremal is taken over expansion points $\xi \in I_{a}\subset I$. We claim that
\[\addtag
\|f(x)-f_N(x)\|\leq \mathcal{M}\bigg|\underbrace{[x-a]\otimes\cdots \otimes[x-a]}_{N+1\text{ copies}} \bigg|; \forall x\in I_a
\]

\begin{theorem}\label{theorm: MMTensor} \textit{
Suppose $H: \mathbb{C}^{\ell} \rightarrow \mathbb{C}^{m\times p}$ is $N+1$ times differentiable on a set $I\subset \mathbb{C}^{\ell}$ that contains vectors $(s_0,p_0)\in (\mathcal{S}\times \mathcal{P}) \subseteq \mathbb{C}^{\ell}$ in its interior. Let $I_{(s_0,p_0)} \subset I $ be a closed and bounded subset of $I$. Then
\begin{align}  \label{eq: UpperMultiSISO}
\|H(s,p)-H_N(s,p)\|\leq \mathcal{M}\Bigg|\underbrace{ \bigg[{s-s_0\atop p-p_0}\bigg]\otimes\cdots \otimes\bigg[{s-s_0\atop p-p_0}\bigg]}_{N+1} \Bigg|\nonumber \\
\end{align}
holds for all $(s,p)\in I_{(s_0,p_0)}$, where
\[\addtag
\mathcal{M}: = \sup_{(\xi_{s},\xi_{p}) \in I_{(s_0,p_0)}}\bigg\|\frac{D^{(N+1)}H(\xi_s,\xi_p)}{(N+1)!}\bigg\|
\]
is the supremum over the induced norm of the tensor $[D^{|n+1|}H]$ over $I_{(s_0,p_0)}$.
}\end{theorem}
Hence, (\ref{eq: UpperMultiSISO}) is the error bound for multivariable function for MIMO system.

The function $H(s,p)$ and $H_N(s,p)$ in (\ref{eq: UpperMultiSISO}) can be interpreted as the (\ref{equ: essTF}) and (\ref{equ: rDSSTF})  at points $(s_0,p_0)$ and $N$th-order reduced model at points $(s_0,p_0)$, respectively. With the desired reduced order $N$ and expansion points $(s_0,p_0)$, the \textit{Theorem \ref{theorm: MMTensor}} gives a prior upper bound on the reduction error using multi-parameter and multi-frequency moment matching method. Furthermore, this error bound allows for finding the optimal expansion points and truncated order.

\section{Example: A thermal parameterized model}
\begin{figure}[h]
  \centering
  \includegraphics[scale=0.11]{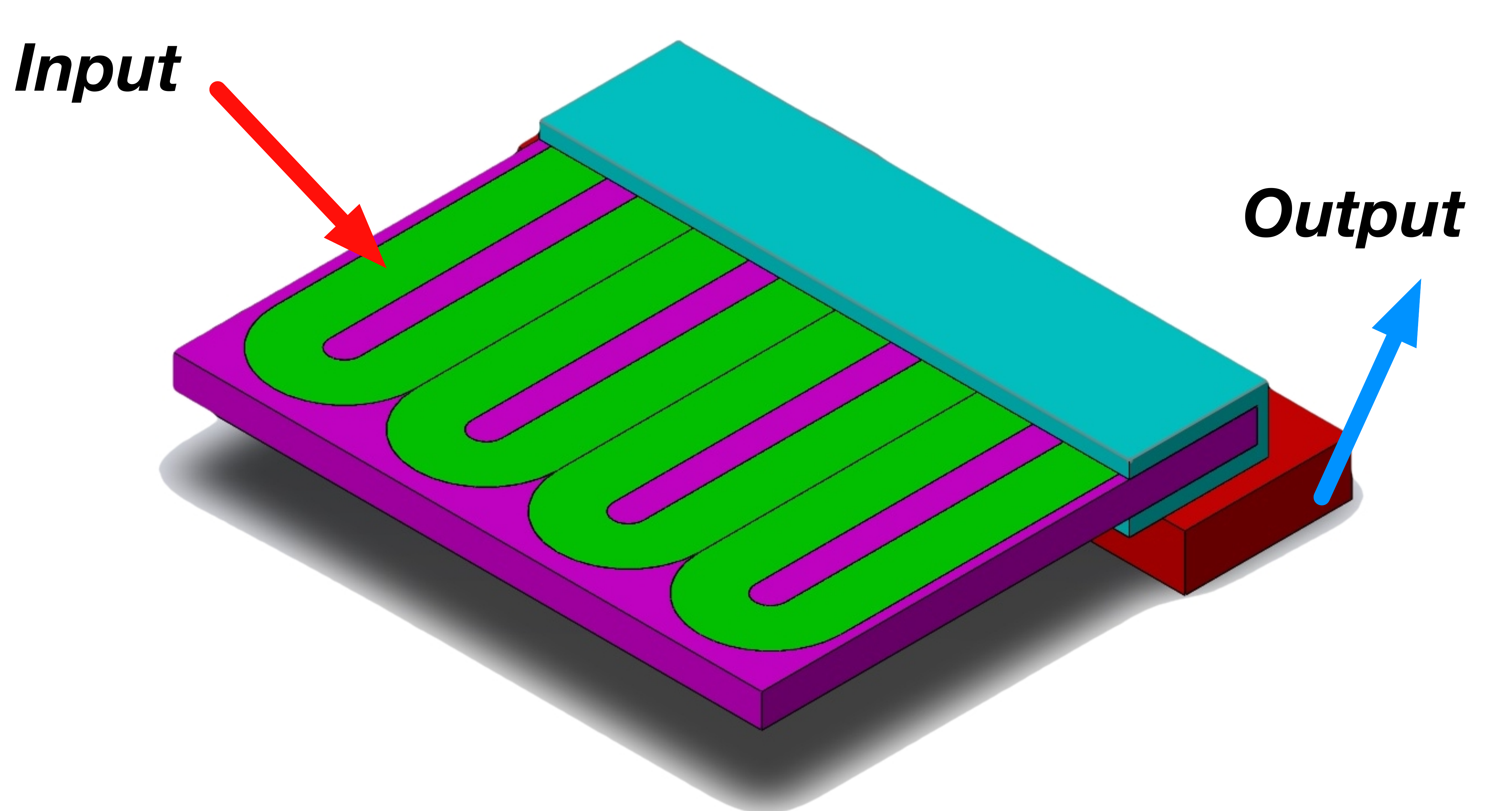}
  \caption{Linear motor with inputs and outputs}\label{Fig:example}
\end{figure}

In this section, we demonstrate some numerical results of the proposed method.
To illustrate the procedure and technique, we consider the example of a linear motor (see Figure \ref{Fig:example}).
The coils of the linear motor generate a magnetic flux that operate along the track. Meanwhile, the coil also creates a heat flux which influences the temperature and, with that the stability of the thermal system. The aim is to predict the transient thermal behavior of the linear motor. Let $\Omega \subset \mathbb{R}^3$ be the spatial configuration space of the motor. Consider the heat equation on $\Omega$ as given by
\begin{eqnarray}
  c_p \rho \cdot \dot{T} &=& \Delta \cdot (\kappa \Delta T) + {q} \quad\text{in} \hspace{0.2cm}\Omega \\
  T(0)& = & T_0 \quad \text{on}\hspace{0.2cm} \partial \Omega_1
\end{eqnarray}
where $T$ is the spatial-temporal dependent temperature of (i.e. $T: \Omega \times [0,\infty) \rightarrow \mathbb{R}$ ) and where the initial temperature  $T_0(w):= T(w,0)$ is given on the boundary $w \in \partial \Omega$ of $\Omega$.  $c_p$ is the material specific heat capacity, $\rho$ denotes the material density and $\kappa$ describes the thermal conductivity. The heat flux $q: \Omega \rightarrow \mathbb{R}$ is spatially distributed and is considered as the input of the system.


Using a finite element (FE) discretization of $\Omega$ and parameterizing the material properties, a parameter-dependent system is generated 
\begin{displaymath}\addtag \label{eq: Example1}
	  			\left\{ \begin{array}{ll}
	  				E({p})\dot{T} =  AT+Bu \\
		    			\hspace{0.8cm}	   y =  CT	\end{array}\right.,
		 	\end{displaymath}
where, with some abuse of notation, $T(t) \in \mathcal{R}^n, n = 1560$ and where $p$ denotes the heat capacity of the material and it is characterized by an affine function $E({p}) =  E_{0}+C_{p}E_{1}$. The state $T$ represents spatial-temporal information which includes $T(t) =\text{col}[T(x_{1},t),\cdots,T(x_{1560},t)]$.

The design goal for this linear motor is to find a suitable material which satisfies the required operating temperature. Therefore, we illustrate three different heat capacities and the $C_p$ varies from $200[J/kg\cdot K]$ to $900[J/kg\cdot K]$. Meanwhile, the operating frequency range of the thermal behavior is mainly located at low frequency (near $1\sim10$ Hz). Under these conditions thereupon, we choose frequencies $s=[0, 1 \pm i]$ and parameters $p =[200,500,900]$.

\begin{figure}[h]
  \centering
  \includegraphics[scale=0.38]{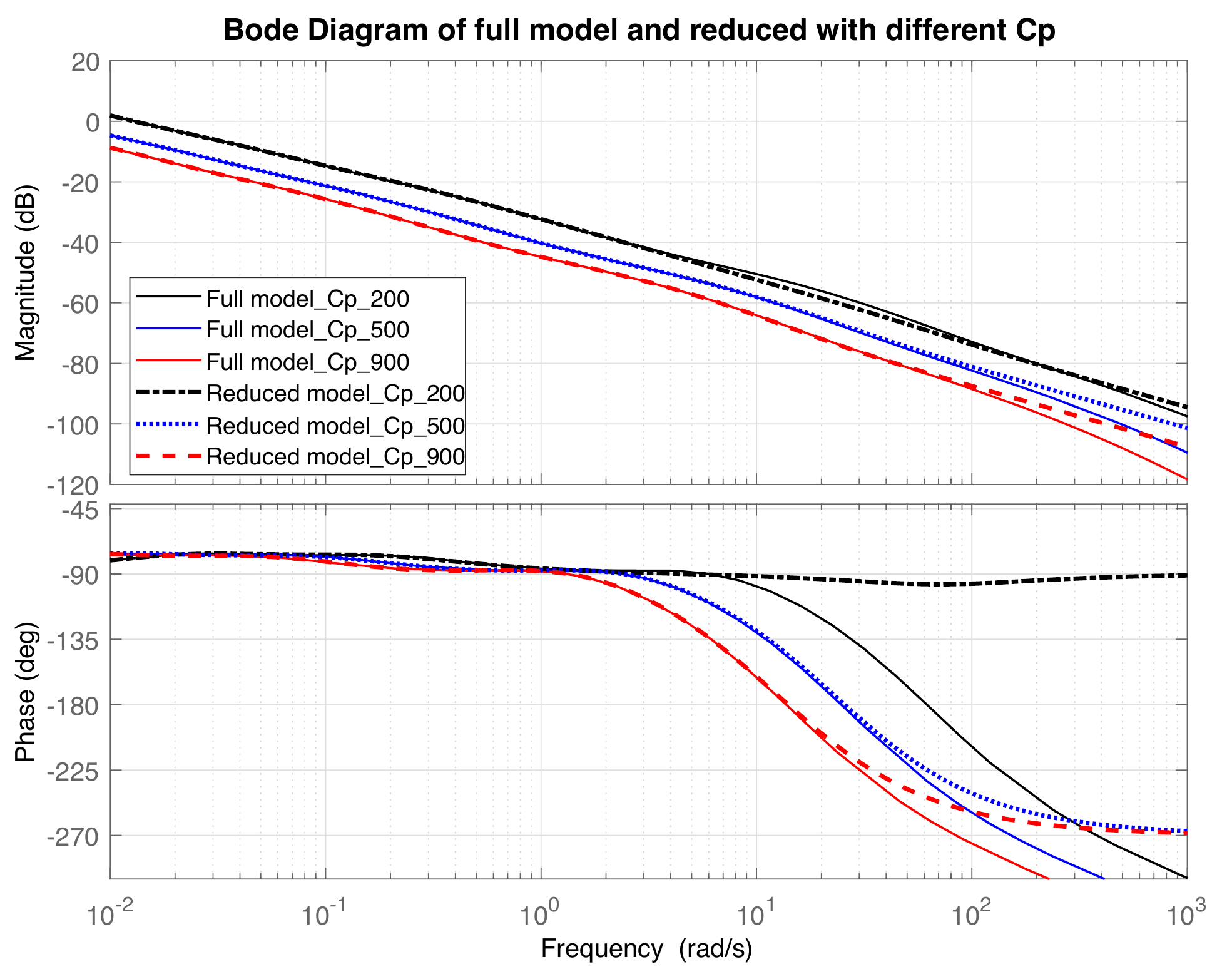}
  \caption{Bode diagram of full model and reduced model with  $C_p=[200,500,900][J/kg\cdot K]$.}\label{Fig:example2}
\end{figure}

Following the procedure in Section III, 
we obtain the reduced model with $r = 12$ orders which includes $6$ expansion points $[0,200],[0,500],[0,900], [1\pm i, 200],[1\pm i, 500],[1\pm i, 900]$. We preserve two moments for each point and obtain the reduced model as \begin{displaymath}\addtag \label{eq: ReExample1}
	  			\left\{ \begin{array}{ll}
	  				E_r({p})\dot{x}_r =  A_rx_r+B_ru \\
		    			\hspace{1.04cm}	   y =  C_rx_r	\end{array}\right.,\quad \text{for } x_r \in\mathbb{R}^{12}
		 	\end{displaymath}
here $E_r(p) = E_{r,0}+C_pE_{r,1}$. As demonstrated in Figure (\ref{Fig:example2}), the solid lines represent the full model with different values of $C_p$. 
The dashed lines denote the reduced order model with corresponding $p$. The frequency range $[0.1, 10]$ in the Bode plot for all three models are matched, and there are some mismatches from $20$ Hz.

\begin{figure}[h]
  \centering
  \includegraphics[scale=0.28]{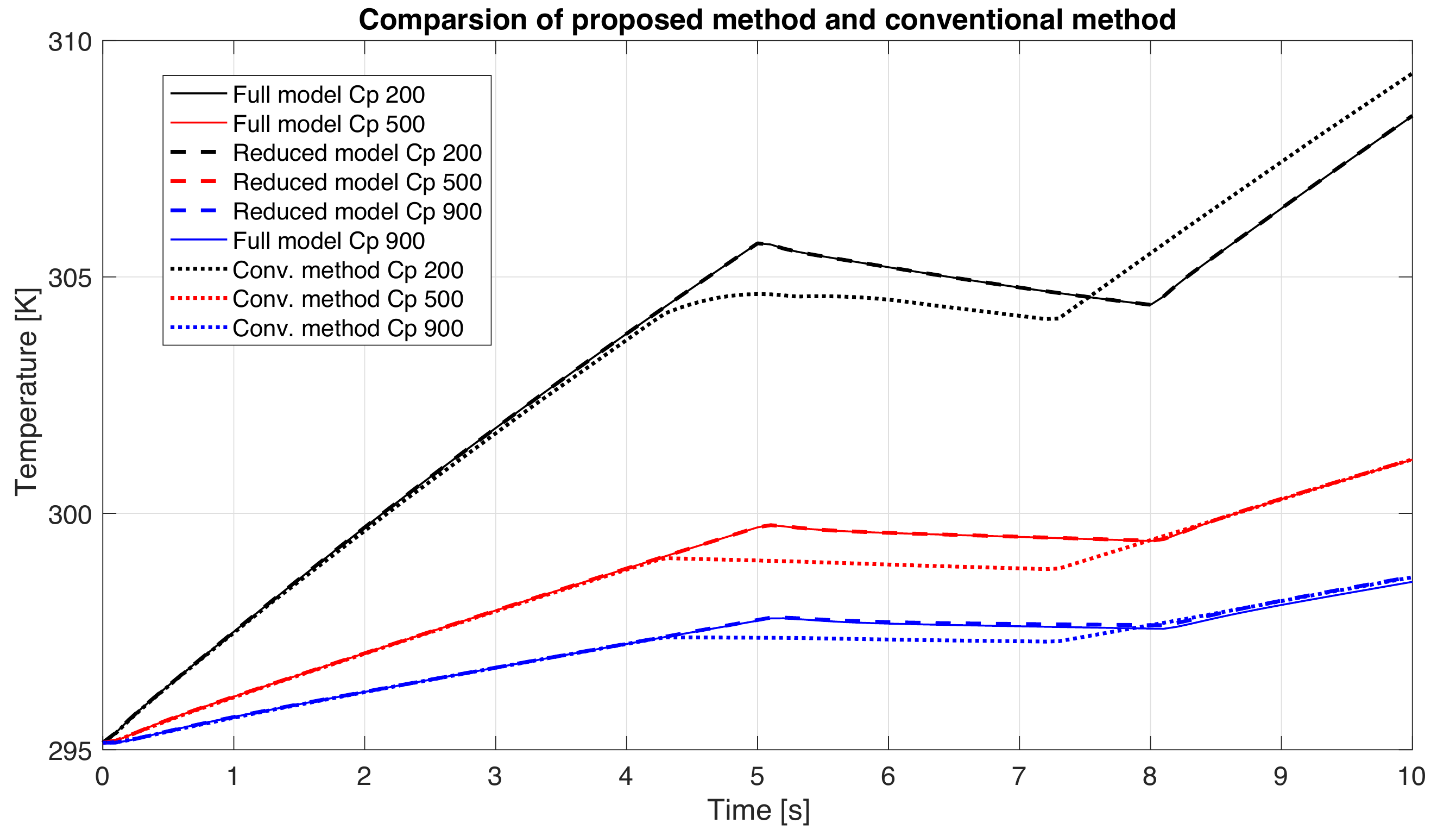}
  \caption{Comparison of full model and reduced model with different heat capacities $C_p=[200,500,900][J/kg\cdot K]$ in time domain simulation.}\label{Fig:example3}
\end{figure}

To show the benefits of the separation of frequencies and parameters in the construction of the projection matrices $V$ and $W$, we also provide a comparison of time domain simulations for the proposed method and the method in \cite{PMORMM}, the COMSOL simulation results which are considered as reference are also included. By applying the same input, the Figure (\ref{Fig:example3}) shows the outputs of all $9$ models with different heat capacities.

In Figure (\ref{Fig:example3}), the solid lines represent the reference of all models with different $p$. The dash lines are the results of the proposed method, the dot lines are the methods from previous work (we choose the same expansion points and same number of moments as the proposed method). Obviously, the proposed method has the better performance which is very close to the reference. The previous method also can match the most of the thermal behavior except for the transient parts. However, the simulation we perform here is only $10$ seconds, it has been observed that the longer simulation time and complicated input are chosen, the performance of the proposed method will not be as good as this one with the same expansion points. Therefore, it leads to another challenging topic which has been left without enough attention, how to choose the optimal expansion points with or without   prior knowledge.

\section{Conclusion}
This paper considers the moment matching problem for linear parameterized system. An explicit algorithm is proposed that establishes a perfect match of moments in both frequency points $\mathcal{S}$ and parameter samples $\mathcal{P}$. An error bound has been derived that represents the local accuracy of the transfer function of the reduced order model nearby the points $\mathcal{S}\times \mathcal{P}$. Simulation results from both time domain and frequency domain show that the proposed method delivers good matching and outperforms the previous work.



\bibliographystyle{abbrv}
\bibliography{libDaming}

\end{document}